\documentclass[a4paper,fleqn]{cas-dc}

\usepackage[authoryear]{natbib}

\usepackage{defs}
\begin{document}
\let\WriteBookmarks\relax
\def\floatpagepagefraction{1}
\def\textpagefraction{.001}

\shorttitle{The impact of jets on the ISM}    

\shortauthors{M. Polletta et al.}  

\title [mode = title]{AGN Feedback: The impact of galactic-scale radio jets on the interstellar medium in starbursting obscured AGN}

\author[1]{Maria del Carmen Polletta}[orcid=0000-0001-7411-5386]
\cormark[1]
\ead{maria.polletta@inaf.it}
\credit{Writing - original draft, Conceptualization, Methodology}
\affiliation[1]{organization={INAF – Istituto di Astrofisica Spaziale e Fisica Cosmica Milano},
            addressline={Via A. Corti 12}, 
            city={Milan},
            postcode={20133}, 
            state={MI},
            country={Italy}}

\author[2]{Carol J. Lonsdale}%[orcid=0000-0003-0898-406X]
%\ead{caroljeanl@mac.com}
\credit{Data curation, Investigation}
\affiliation[2]{organization={Associated Universities, Inc.},
            addressline={6750 Park Tower Dr}, 
            city={Vienna},
            postcode={22180}, 
            state={VA},
            country={USA}}

\author[3]{Patil Pallavi}%[orcid=0000-0002-9471-8499]
%\ead{ppatil13@jh.edu}
\credit{Data curation, Investigation}
\affiliation[3]{organization={William H. Miller III Department of Physics and Astronomy, Johns Hopkins University},
            addressline={3400 N. Charles Street}, 
            city={Baltimore},
            postcode={21218}, 
            state={MD},
            country={USA}}

\author[1]{Giustina Vietri}%[orcid=0000-0001-9155-8875]
%\ead{giustina.vietri@inaf.it}
\credit{Data curation, Methodology}

\author[2]{Amy E. Kimball}%[orcid=0000-0001-9324-6787]
%\ead{akimball@nrao.edu}
\credit{Data curation, Investigation}

\author[1]{Paolo Franzetti}%[orcid=0000-0002-6986-0127]
%\ead{paolo.franzetti@inaf.it}
\credit{ETC developer}

\author[4]{Colin J. Lonsdale}%[orcid=0000-0003-4062-4654]
%\ead{cjl@mit.edu}
\credit{Data curation, Investigation}
\affiliation[4]{organization={MIT Haystack Observatory},
            addressline={99 Millstone Road},
            city={Westford},
            postcode={01886}, 
            state={MA},
            country={USA}}
            
\cortext[1]{Corresponding author:}

\begin{abstract}
A highly star-forming galaxy at $z\sim2$ hosting an obscured, luminous
active galactic nucleus (AGN) and a relativistic radio jet sets the stage
for a cosmic crime scene.  The victim is star formation, the suspect
is AGN feedback.  These systems offer a rare opportunity to catch this
process in the act. We propose SHARP@ELT integral-field spectroscopic
observations of heavily obscured, luminous AGN with resolved radio emission
to witness the onset of feedback and its impact on the host interstellar
medium (ISM).  Such objects trace a short-lived ($<10^5$\,yr)
evolutionary phase in which a recent starburst, newly triggered radio jets,
and a deeply embedded AGN coexist.  In this phase, feedback is expected to
suppress star formation while clearing the dusty nuclear regions.  Using
SHARP/VESPER, we will derive spatially resolved maps of stellar ages, star
formation rate density, gas density, and ionization state, alongside the
kinematics of stars and ionized gas.  By directly comparing these properties
with the radio structures, we will quantify the effects of both jet-driven
and radiative feedback on the ISM.  Our sample consists of eleven luminous,
obscured AGN at $1.5<z<2.5$ with resolved radio emission on scales of
$\sim$1–-15\,kpc.  Exploiting the VESPER multi-Integral Field Selector
capability, we will obtain resolved continuum and emission-line maps for at
least 110 galaxies at cosmic noon, enabling a comprehensive characterization
of their environments, multiphase ISM, and nuclear activity within 55 hours
of integration time.  SHARP will thus reveal AGN feedback at the epoch when
it is most effective, providing a decisive step toward understanding its
role in galaxy evolution.
\end{abstract}

% Keywords

\maketitle

% Main text
\section{Scientific Rationale}\label{sec:intro}

\begin{figure}%[ht]
    \centering
    \includegraphics[width=0.5\textwidth]{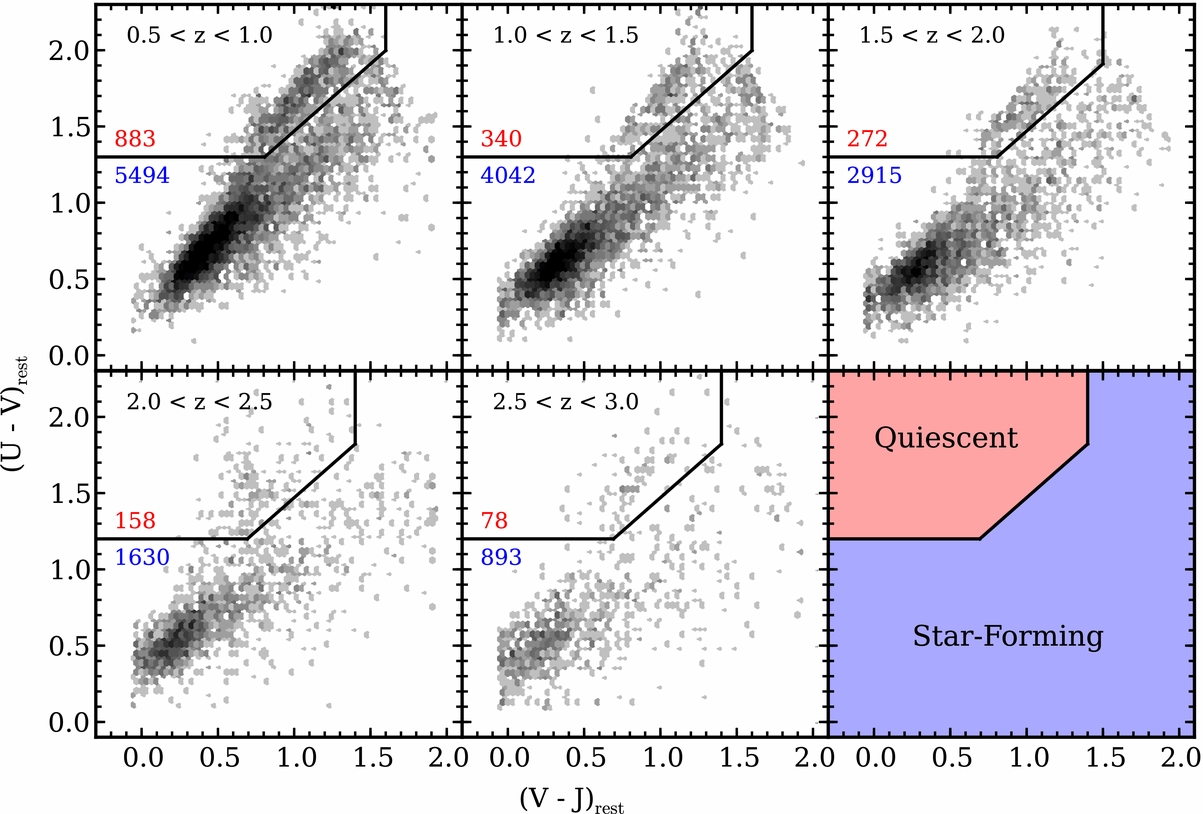}
   \caption{Rest-frame UVJ diagrams used to separate star-forming and quiescent galaxies as indicated in the bottom-right panel. All galaxies with masses $>10^{9.5}$\msun\ from the ZFOURGE survey are shown \citep[from][]{tomczak14}.}
    \label{fig:uvj_diagram}
\end{figure}

\begin{figure*}%[ht]
    \centering
    \includegraphics[width=0.95\textwidth]{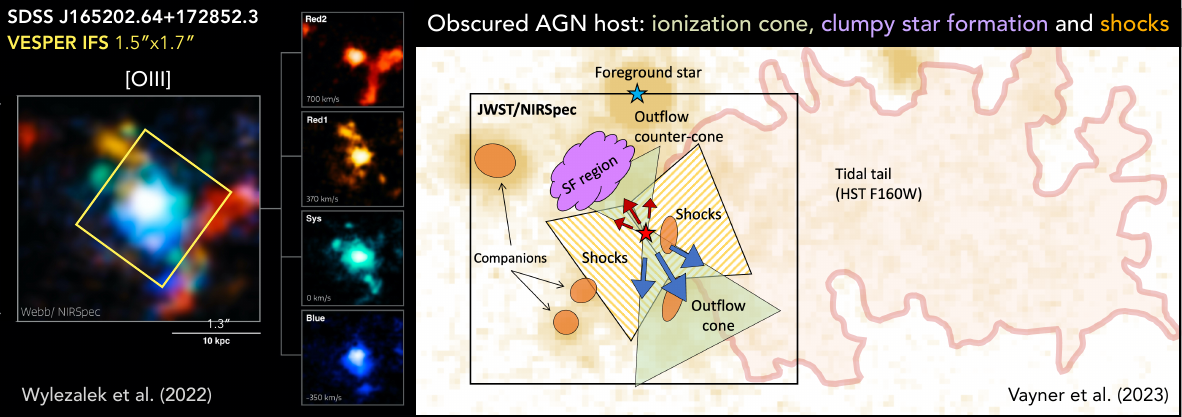}
   \caption{JWST/NIRSpec 3\arcsec$\times$3\arcsec\ image of the obscured AGN J1652 showing complex \oiiib\ kinematics 
   and morphology (left panel). The yellow rectangle shows the FOV of a single VESPER IFS. The four middle panels show intensity maps of the \oiiib\ emission at velocities $-$350, 0, 370, and 700\,\kms\ and width of $\sim$120\,\kms\ \citep{wylezalek22}. The right panel shows an illustration of the different kinematic components detected in this object, including some companion galaxies and a tidal tail \citep{vayner23}. }
    \label{fig:red_QSO}
\end{figure*}

Active galactic nuclei (AGN) play a major role in the evolution of galaxies and galaxy clusters \citep{dimatteo05,dave19,fabian12}.  The energy they release influences the surrounding environment across a wide range of physical scales, making an understanding of these interactions essential for explaining many phenomena observed in galaxies and clusters.  This interaction, commonly referred to as AGN feedback, is a key component of cosmological and semi-analytical simulations, although implementation methods differ across simulations as many of the underlying physical processes remain only partially constrained \citep{lagos25,maragkakis25}.\\
AGN feedback is typically modeled either as kinetic energy injection through jets \citep[radio mode, ][]{silk98,dave19,kurinchi_Vendhan24} or as an AGN radiation-driven wind or outflow \citep[quasar mode, ][]{dimatteo05,hopkins08}. Radio mode is principally invoked to prevent gas cooling or to maintain the quiescence of massive galaxies \citep[maintenance mode feedback; ][]{mcnamara12} in which powerful jets can “drill” through a galaxy interstellar medium (ISM), reach the intergalactic medium (IGM) and affect the surrounding environment on large scales. On the other hand, radio jets of low-power are expected to impact the host ISM, drive shocks over a wide solid angle and deposit their energy more broadly than powerful jets \citep{mukherjee25}.  These sub-galactic scales (100s of pc to a few kpc) radio jets might be recently triggered \citep{reynolds97,jin25} or the result of propagation through a particularly dense ISM, where jets are impeded from expanding \citep{stanghellini25}.  In this case, the jets can stir turbulence within the ISM, potentially reducing star formation \citep{mandal21,leftley24}.  Under certain circumstances, they might instead trigger star formation \citep[positive feedback;][]{santoro18,capetti22}. \\
The effects of jets have been relatively well studied in large samples of AGN at low redshift ($z<0.5$) and in a few radio-loud AGN with galaxy-scale jets at high redshifts \citep[$z>1.5$; see][ for a comprehensive review]{mukherjee25}. Significant progress in our understanding of the jet-host interaction at low redshifts has been enabled by integral field spectroscopic observations \citep[see e.g.,  the GATOS and the QSOFEED surveys; ][]{garcia_bernete24,zhang24_gatos,davies24,bessiere24}.  These studies find that radio jets interact with the ISM, might generate turbulence and drive ionized outflows.  For example, \citet{ulivi24} report a strong enhancement in the emission-line velocity dispersion perpendicular to the direction of the radio jet and a correlation between the gas mass and energetics and the radio power.  The role played by jets of modest power as dominant feedback mechanism has been also confirmed in several radio quiet luminous AGN \citep{zakamska14,villar_martin17,kellermann16,jarvis19,roy24,cresci23,harrison24,gingras25}.\\
Little is known about the interplay of small-scale jets and their host at intermediate and high redshifts. This is because identifying galaxies that host compact radio jets is observationally demanding with current facilities and the interaction might occur deep within a heavily obscured core  \citep{hopkins06b}.  Furthermore, this type of investigation requires observations that trace the stellar and gas kinematics at a spatial resolution comparable with that of the radio images (sub-arcsec resolution). The few studies of the radio-ISM connection carried out in objects with moderate-to-high radio luminosities at high redshifts find evidence of a strong connection between the radio emission and ionized outflows \citep[e.g.,][]{hwang18,ilha25,he25}, but the origin of the radio emission is debated. Indeed, it is often not clear whether it is associated with a relativistic jet or with electrons accelerated by an outflow-driven shock.\\
To understand the different feedback modes, we need to spatially and kinematically separate winds, outflows, and radio jets, quantify their energetics and assess their impact on the ISM. Since both the global SFR density and the BH accretion rate (BHAR) start declining at $z\lesssim2$ \citep{madau14,kim24}, AGN feedback is expected to be the most effective around this epoch for the bulk of the galaxy population.  This is supported by the increase in the number of quiescent galaxies at $z<2$, as illustrated in Fig.~\ref{fig:uvj_diagram} \citep[from ][]{tomczak14}. 
Thus, the ideal systems for such a study are  luminous and obscured AGN on the verge of shutting down and with a radio jet capable of causing such a transition. Such a study requires high resolution maps that trace the distribution and the kinematics of the stellar population, of the ISM and of the radio jet. \\
An example of the complex system we expect to find is illustrated by the luminous and obscured AGN SDSS\,J165202 $+$172852 at $z\sim2.9$ (J1652, hereinafter). In Fig.~\ref{fig:red_QSO}, we show a James Webb Space Telescope (JWST)/NIRSpec map of the \oiiib\ emission and a scheme of the multiple kinematic components detected in this system: a radiation-driven outflow, shocked gas perpendicular to the outflow and star forming clumps \citep{wylezalek22,vayner23}. Since the radio emission is unresolved in the available observations ($\gtrsim1.2$\arcsec\ resolution), it is not clear whether it is due to shocks induced by the outflow, or to an AGN-driven radio jet \citep{hwang18}. An additional element to consider in the analysis of this kind of system, it is the environment. For example, this AGN seems to lie in the core of a galaxy protocluster because several interacting companions are detected within 12\,kpc. \\
The SHARP instrument \citep{sharp24} designed for the Extremely Large Telescope (ELT) offers the spectral coverage and resolution, sensitivity and spatial resolution to carry out this type of study and investigate the role of the radio jet if high resolution  radio images are also available. With its capabilities, it will be possible to trace the effects of jets and winds by mapping the kinematics of the stellar population and of the ionized gas in the host galaxy to scales of a few hundreds of parsecs, and overcoming the effects of dust extinction.

\section{AGN feedback revealed by SHARP}

The main goal of the scientific case presented here is to understand the impact of small scale jets, and of AGN radiation in star-forming AGN at cosmic noon. Sub-arcsec resolution maps at radio and visible wavelengths are necessary to trace their stellar populations, the warm and cold gas, and the jet. Resolved spectral measurements at rest-frame visible wavelengths are required to disentangle different energetic sources (AGN, star formation, shocks), reveal rotating, outflowing and inflowing gas, and locate star forming regions with scales of a few hundreds parsec deep into the nuclear regions. With the ELT spatial resolution and sensitivity and the SHARP spectral coverage ($\lambda=1.2-2.4\mu$m) and resolution (R$\sim$3000) it will be possible to obtain these types of maps and kinematic measurements for AGN at cosmic noon.  These measurements will permit to assess the role of different feedback mechanisms when they were the most impactful.

\subsection{Selected targets: the radio DOGs}

We have identified as ideal targets for the proposed study the so-called radio dusty obscured galaxies (DOGs).  These galaxies host a luminous and heavily obscured AGN with a sub-galactic scale radio jet and are actively forming stars.  According to the current AGN evolutionary picture, the growth of SMBHs initially occurs enshrouded in central dust that was funneled inward along with gas, triggering an obscured starburst and rapid AGN fueling. The radio DOGs were selected to possibly represent this growing phase, and their radio properties make them good candidates to witness the onset of radio mode feedback.
\begin{figure*}%[ht!]
    \centering
    \includegraphics[width=0.22\textwidth]{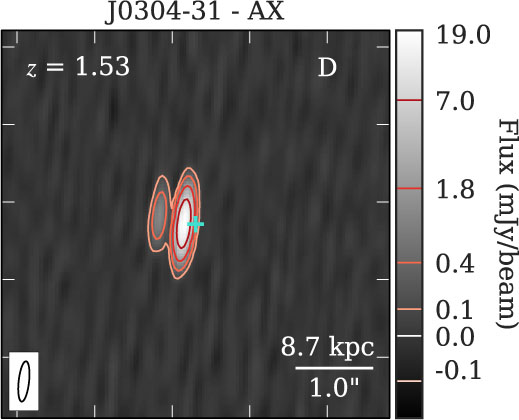}
    \includegraphics[width=0.22\textwidth]{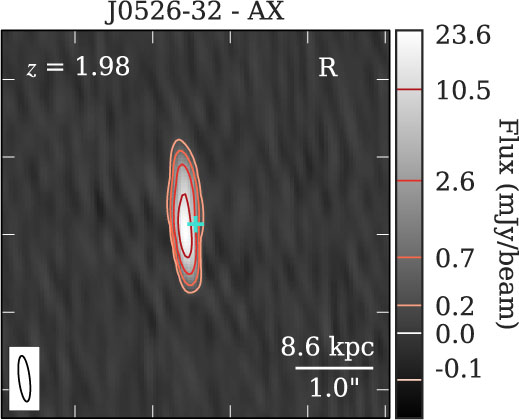}
    \includegraphics[width=0.22\textwidth]{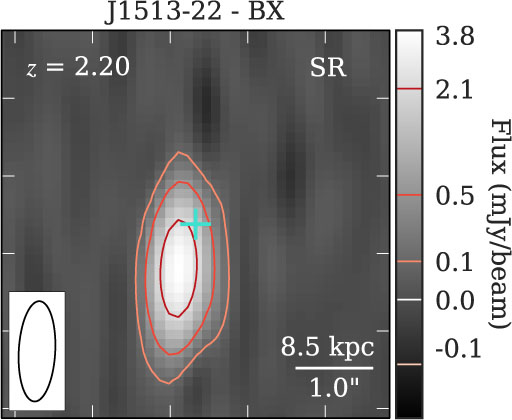}
    \includegraphics[width=0.22\textwidth]{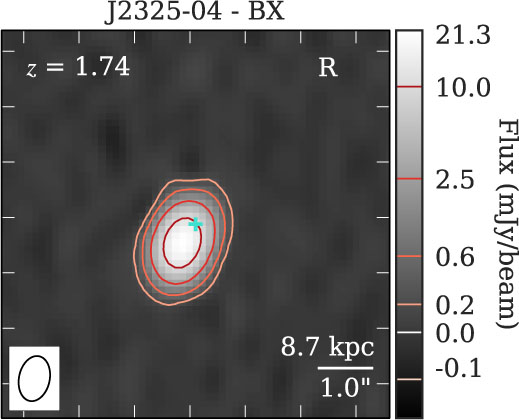}
   \caption{VLA 10\,GHz continuum images of four radio DOGs from the selected sample (see Table~\ref{tab:targets}). The red contours represent the radio flux starting from 5$\sigma$ with steps of 4$\sigma$ (see colorbar to the right). The cyan plus sign is the WISE position. The VLA synthesized beam is shown as a black ellipse in the lower-left corner. The source name, redshift and radio morphology of each source are annotated. \citep[Adapted from][]{patil22}.}
    \label{fig:vla_imgs}
\end{figure*}
\begin{figure}%[ht!]
    \centering
    \includegraphics[width=0.45\textwidth]{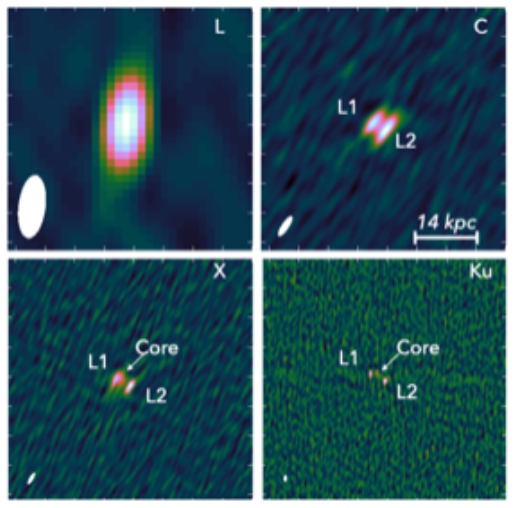}
   \caption{VLA images at different frequencies (L: 1.5\,GHz, C: 6\,GHz, X: 10\,GHz, Ku: 15\,GHz) and angular resolution ($\sim0.1-2$\arcsec) of a radio DOGs. The synthesized beam is shown as a white ellipse in the lower-left corner in each panel.}
    \label{fig:multi_freq_imgs}
\end{figure}

\begin{figure*}%[ht]
    \centering
    \includegraphics[width=0.95\textwidth]{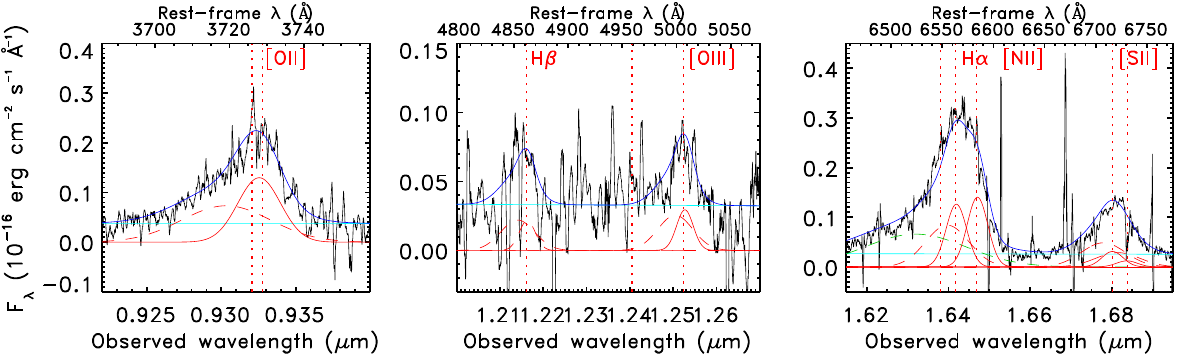}
   \caption{Visible emission lines (\oii\ in the left panel, \hbeta\ and \oiiib\ in the middle panel, and \Ha, \nii, and \sii\ in the right panel) from an X-Shooter spectrum \citep[solid black line; ][]{ferris21} of the radio DOG J1500$-$06. The visible lines are modeled with two gaussian components, one (red solid lines) at $z=1.5015$ (expected observed wavelengths are shown as the vertical dotted lines), and one blueshifted by $\sim-$360\,\kms\ (red dashed lines).  The modeled continuum (cyan line) and the sum of the modeled lines and continuum (blue line) are also shown.  The large \oii/\oiiib\ ratio indicates the presence of shocks and the blueshifted components indicate the presence of outflowing gas.}
    \label{fig:J1500_spectrum}
\end{figure*}

\begin{figure}%[ht!]
    \centering
    \includegraphics[width=0.5\textwidth]{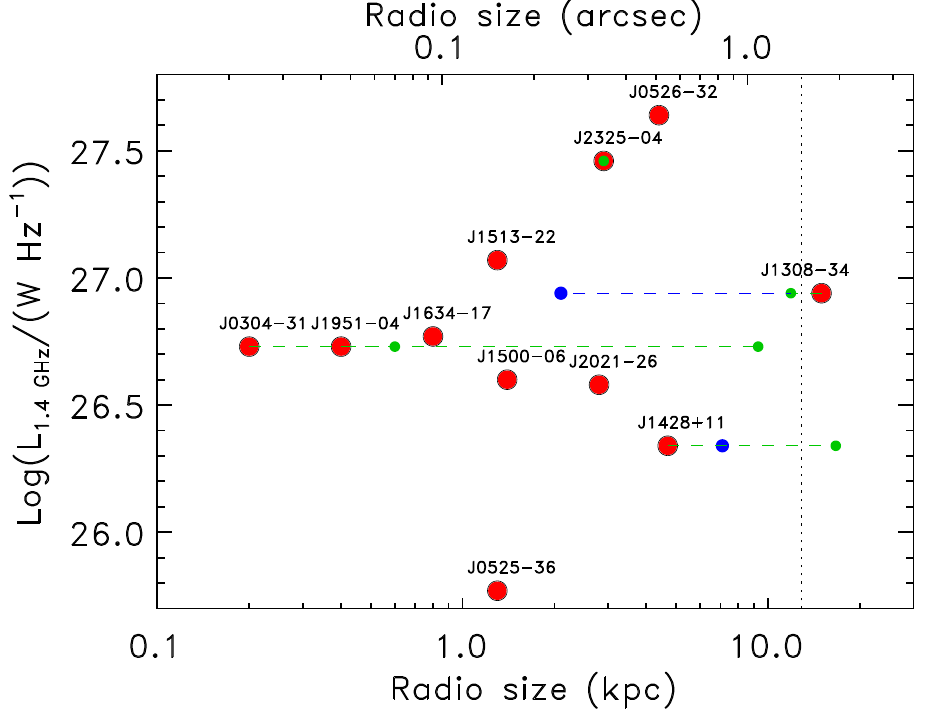}
   \caption{Radio luminosity at 1.4\,GHz and radio source size (major axis) for each detected component (red, and green circles) of the selected radio DOGs. The sources are labeled with their name. The radio source sizes are given in arcsec on the top axis and in kpc on the bottom axis assuming $z=2$. The dotted vertical line represents the size of a single VESPER IFS (1.5\arcsec). In all, but one case a single IFS covers the single radio components, but multiple IFSs are necessary to cover the full extent of complex radio structures with multiple components.}
    \label{fig:lrad_size}
\end{figure}

These radio DOGs were selected as extremely red and bright mid-infrared (MIR) sources using the WISE all sky survey and as unresolved radio sources in the NVSS (45\arcsec\ beam) and FIRST ($\sim$5\arcsec\ beam) radio surveys \citep{lonsdale15}.  The initial sample contains 167 radio DOGs, all are obscured AGN at $z=0.5-3$.  Follow-up multi-frequency (150\,MHz-–10\,GHz) radio imaging with the JVLA, and the VLBA at various spatial resolutions (0.065\arcsec–2\arcsec) revealed sub-galactic jets with intermediate power and peaked spectra for 72\% of the sample \citep{patil20,patil22}, consistent with compact radio jets \citep{odea21} (see some examples in Fig.~\ref{fig:vla_imgs}, and~\ref{fig:multi_freq_imgs}).  Their high radio luminosity and morphologies (see Figs.~\ref{fig:lrad_size}, and~\ref{fig:vla_imgs}) imply the presence of a jet that has been recently launched.
Many jets show bends and complex morphologies as seen in simulations of moderate power jets driving into a dense ISM.  Visible and IR imaging, sub-mm/mm data and spectroscopic observations are available for 49 radio DOGs providing, through spectral energy distribution (SED) modeling \citep[carried out with both \texttt{CIGALE} and radiative transfer models;][]{boquien19,efstathiou00}, stellar masses, SFRs and AGN fractions \citep{lonsdale15}.  About half of these sources are well above ($>0.5$\,dex) the star forming main sequence \citep[MS;][]{popesso23} and are thus considered starbursting.  Many of these systems also exhibit high-excitation visible emission lines \citep[Fig.~\ref{fig:J1500_spectrum}; ][]{kim13,lonsdale15,ferris21}, implying that they are also radiatively efficient AGN.  In at least one case, J1500$-$06, there is evidence that either a wind or the jet is driving the ionised gas outward and generating shocks.  This interpretation is supported by spectroscopic observations of J1500$-$06 \citep[see Fig.~\ref{fig:J1500_spectrum}; ][]{ferris21}, which reveal a blueshifted component in several visible emission lines (\oii, \Ha, \sii, and \oiiib) as well as unusual visible line flux ratios (e.g., high [OII]/[OIII]).  \\
There are eleven radio DOGs visible from Cerro Amazones ($-70$\deg$<\delta<20$\deg), with resolved (in images with 0.1\arcsec\ resolution) sub-galactic or slightly larger scales radio jets (up to 17\,kpc), and with spectroscopic redshift between 1.5 and 2.5 (excluding the 1.7--1.98 range).  This redshift range assures the observability of major spectral features such as \oiiib, and \Ha\ in regions of high sky transmission.  The sources' name and main properties are listed in Table~\ref{tab:targets}. The selected radio DOGs represent a rare phase where intense star formation and a radio jet have been recently triggered in a heavily obscured core. In a few cases, multiple radio components are observed (see sources with double or triple radio morphology in Table~\ref{tab:targets}). These might represent a jet that has recently broken out the host ISM or multiple phases of moderate radio activity on short timescales ($<10^5$\,yrs). Even though this sample is quite small, it covers a broad range of jet sizes that might be the results of different ISM conditions, jet ages and energetics. They are thus the ideal targets to witness the onset of feedback, the first signs of clearing out the central dust, and to study the interaction between the ISM and the AGN radiation and radio jet.

\begin{table*}[ht!]
\caption{Selected radio DOGs and main properties}\label{tab:targets}
\begin{tabular*}{\tblwidth}{@{}CCC CCC CCC C@{}}
\toprule
     Target   & $\alpha$     & $\delta$        & Redshift & K mag & Radio & Radio & Log(L$_{\rm 1.4GHz}$) &   SFR$^b$    & M$_{\rm star}^b$ \\
     name     & (h:m:s)  & (\deg:\arcmin:\arcsec)&        & (AB)  & Morph$^a$ & Size (kpc)   &  (W\,Hz$^{-1}$)     & (\msun/yr$^{-1}$) & (10$^{10}$\msun)  \\
\midrule
    J0304-31  & 03:04:27.54  & $-$31:08:38.28  &  1.530 &  20.25$^b$ &   D   &  $<$0.2, $<$0.6     &    26.73  &    380$\pm$166    &    5.9$\pm$ 1.5   \\
    J0525-36  & 05:25:33.47  & $-$36:14:40.93  &  1.688 &  21.12$^b$ &   R   &  1.3             &    25.77  &    188$\pm$\070   &    4.9$\pm$ 1.8   \\
    J0526-32  & 05:26:24.72  & $-$32:25:00.87  &  1.980 &  20.35$^b$ &   R   &  4.4             &    27.64  &    688$\pm$\034   &    6.3$\pm$ 0.3   \\
    J1308-34  & 13:08:17.01  & $-$34:47:54.36  &  1.652 &  20.64     &   T   & 15.0, 11.9, 2.1  &    26.94  &    218$\pm$\084   &    6.9$\pm$ 1.9   \\ 
    J1428+11  & 14:28:59.70  &  \011:13:18.79  &  1.600 &  20.00$^b$ &   T   &  4.7, 16.7, 7.1  &    26.34  &   1000$\pm$\050   &    9.2$\pm$ 0.5   \\ 
    J1500-06  & 15:00:48.73  & $-$06:49:39.84  &  1.500 &  20.51     &   R   &  1.4             &    26.60  &    353$\pm$\018   &    3.2$\pm$ 0.2   \\
    J1513-22  & 15:13:10.42  & $-$22:10:04.62  &  2.200 &  20.29     &   SR  &  1.3             &    27.07  &    530$\pm$363    &   30.$\pm$ 16.   \\
    J1634-17  & 16:34:26.87  & $-$17:21:39.48  &  2.070 &  21.62$^b$ &   SR  &  $<$0.8          &    26.77  &    346$\pm$\069   &    3.7$\pm$ 1.0   \\
    J1951-04  & 19:51:41.23  & $-$04:20:24.60  &  1.580 &  20.44     &   D   &  $<$0.4, 9.3     &    26.73  &    181$\pm$104    &   11.$\pm$ 8.   \\ 
    J2021-26  & 20:21:48.06  & $-$26:11:59.29  &  2.440 &  21.80     &   SR  &  2.8             &    26.58  &    483$\pm$307    &   40.$\pm$ 33.   \\
    J2325-04  & 23:25:05.08  & $-$04:29:48.12  &  1.737 &  19.60$^b$ &   R   &  2.9, 2.9        &    27.46  &    420$\pm$249    &   25.$\pm$ 8.   \\
\bottomrule
\end{tabular*}
{\small $^a$Radio morph describes the morphology of the radio emission ( SR: Slightly resolved, R: Resolved and single component, D: Double, T: triple; from \citealt{patil22}).
$^b$ Estimated from SED fitting using \texttt{CIGALE} \citep{boquien19}.}
\end{table*}

\subsection{Measurements} 
\label{sec:strategy}

\begin{figure*}%[ht!]
    \centering
    \includegraphics[width=0.95\textwidth]{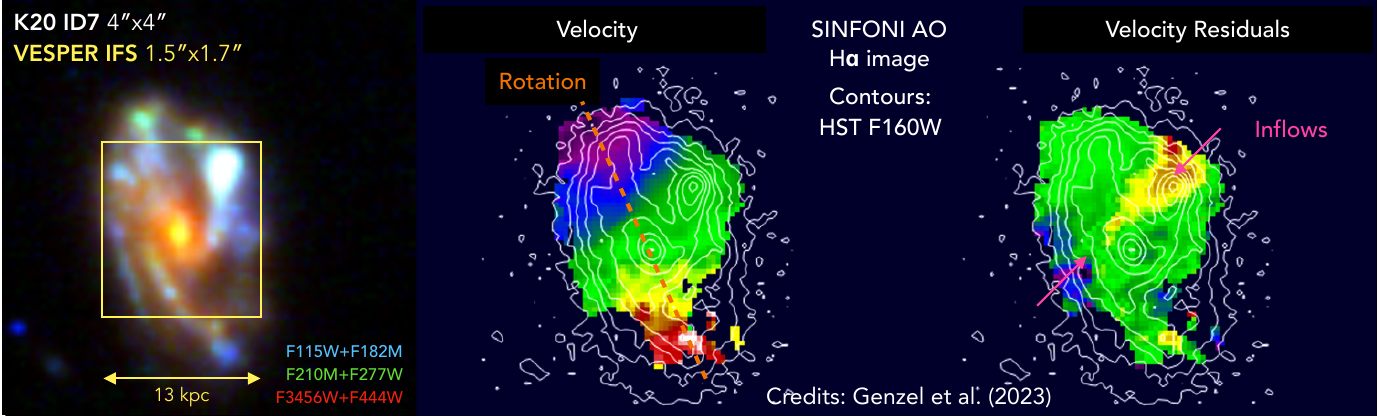}
   \caption{JWST multi-band image of the galaxy K20 ID7 at $z=2.225$
(\citet{tacchella15}; JWST images credits: DAWN JWST Archive;
\citet{grizli,valentino23}; left panel). The yellow rectangle represents the FOV of a VESPER IFS. The middle panel shows the SINFONI \Ha\ velocity map and the right panel the velocity residual map \citep{genzel23}. The white contours are the H-band emission. The dashed orange line represents the galaxy rotation plane and the magenta arrows the inflowing gas. (Adapted from \citet{forster_schreiber24}).}
    \label{fig:K20_ID7}
\end{figure*}

The proposed observational program aims at carrying out (1) a structural analysis of the selected targets to identify their different components (nucleus, jet, bulge, disk, star-forming clumps), (2) a spectral analysis to determine the different sources of ionization (AGN radiation, star formation, radio jet, shocks), (3) a spectrophotometric analysis to build resolved maps of stellar ages, SFR density, stellar mass and extinction, (4) a kinematic analysis of the stellar component, of the warm and cold phases of the ISM, and of the gas non-rotational components, (5) a study of the impact of the jet on the ISM, and (6) an investigation of the environment. Below we describe the measurements we will obtain and diagnostic tools we will apply for each task.\\
\textbf{(1) Galaxy decomposition:} 
We selected targets at redshift $1.5<z<2.5$ in order to observe their rest-frame visible (4800--6800\,\AA) emission with VESPER.  Star-forming galaxies at these redshifts are often rotating disks with star-forming clumps.  In Fig.~\ref{fig:K20_ID7}, we show an example of galaxy at $z\sim2$ imaged with JWST \citep{forster_schreiber24}.  The multi-band image, created using Trilogy \citep{trilogy}, shows color variations across the galaxy that highlight star forming clumps, dusty regions, spiral arms, and faint extended structures.  The complexity of this image illustrates the need of decomposing the integrated light in order to reveal the different mechanisms at play and their origin. The VESPER Integral Field Selector (IFS) spaxel size (31\,mas, equivalent to $\lesssim300$\,pc at these redshifts) is ten times smaller than the expected average effective radius \citep[i.e., $R_{\rm e}\simeq3$\,kpc; ][]{vanderwel25} thus permitting to resolve the galaxy light in hundreds spaxels and separate the different regions such as the nucleus, the jet, the bulge, the disk, star-forming clumps and other sub-structures within a single IFS. Multiple IFS will be employed to cover the most extended targets, as it would be the case for J1652, and K20 ID7 shown in Figs.~\ref{fig:red_QSO}, and~\ref{fig:K20_ID7}.\\
\textbf{(2) Spectral analysis :}
The visible rest-frame spectrum of AGN and star-forming galaxies is rich with spectral features that can be used to constrain the properties of the stellar population, and of the ionized and cold gas in the ISM.  Hydrogen (Balmer series) or metal (Mg, Fe, and others) absorption features can be used to measure the distribution of stellar ages and metal abundances \citep{leitherer10}. The Na\,I\,D absorption feature can be used to search for powerful neutral outflows, common among massive galaxies at cosmic noon \citep{davies24}.  Since all selected galaxies are AGN and star-forming, their spectra should be rich with emission lines from ionized gas, such as \hbeta, \oiiib, \Ha, and \sii. Line ratios, such as \nii/\Ha, and \oiiib/\hbeta, can be used to apply the BPT diagnostic diagram \citep{BPT_diagram},  identify regions with different ionization states and temperatures \citep[with i.e., \oiiib, \hbeta, \oii, \heii4686, \oiii4363; ][]{osterbrock06,liu13}, and build metallicity maps \citep{maiolino19}. The ratio of the trans-auroral emission lines \sii\ will be used to estimate the ionized gas electron density which enters in the mass outflow rate, and in the kinetic power and efficiency.
The \oii\ line, available for the sources at $z>2.2$, is a powerful indicator of compression due to shocks \citep[see e.g., ][]{santoro20}.  The high ionization feature \gion{FeV}{2} $\lambda$5158 can be also used as an indicator of AGN  photoionization and shocks \citep{zakamska14}.  \\
The expected variations in these lines might be subtle, requiring high S/N values ($\gtrsim$10) to measure their fluxes with sufficient accuracy.  In particular, this applies to the weak broad components that vary as a function of kinematics and/or luminosity. The spectral analysis will enable to discriminate between shocks, young stars, and nuclear radiation.\\
\textbf{(3) SED fitting:}
Resolved spectro-photometric fitting will provide detailed maps of stellar ages, SFRs, ionization intensity, and dust extinction.  For this task we will take advantage of the mm and radio data as their inclusion is crucial to break the age-extinction degeneracy and accurately estimate obscured SFRs and stellar ages \citep{li24_cristal}. \\
\textbf{(4) Kinematic analysis:}
VESPER spectral resolution (R$\sim$ 3000) will enable the successful removal of the contribution of sky lines and the identification of sub-components characterized by different velocity shifts, widths, and intensities (see e.g., Fig.~\ref{fig:J1500_spectrum}). 
The gas kinematics will be investigated to characterize the disk dynamical status and identify non-rotational components, such as inflows, and outflows. Examples of the type of analysis that the VESPER data will enable are shown in Fig.~\ref{fig:red_QSO}, and~\ref{fig:K20_ID7}. In case of J1652, different kinematic components of the \oiiib\ emission are identified as a double outflow and shocked gas perpendicular to it, plus a star-forming region associated with the outflow \citep{wylezalek22,vayner23}. In case of K20 ID7, the \Ha\ velocity map shows a rotating disk and the residual map two inflows almost perpendicular to the disk \citep{genzel23,forster_schreiber24}.
We will carry out a similar analysis on the selected targets. The different velocity components will be separated and for each of them we will quantify the spatial extent, and the luminosity. \\
At $z\lesssim2$, the kinematical properties of star-forming galaxies experience a dramatic change, the disk velocity dispersion $\sigma_{\rm disk}$ strongly decreases and the dynamical support, given by the ratio between rotational velocity and $\sigma_{\rm disk}$, rapidly increases, independently on the probed ISM phase \citep[e.g.,][]{lee25}. This trend has been attributed to decreasing gas fraction at lower redshifts \citep{tacconi20}. Such a decrease is likely linked to the decline in BH accretion and SFR, probably a consequence of AGN feedback. Our targets straddle this epoch and might thus provide observational evidence for this change in the disk dynamics and indicate the cause.\\
Resolved kinematic maps can be used to identify regions with variations in velocity dispersion and turbulence and to investigate their origin. The power of this kind of study is demonstrated by the JWST GARDEN survey (PI: S. Kassin, Program ID: 2123) where kinematics maps of \Ha\ in $z\sim2$ galaxies reveal higher velocity dispersions in the inter-arm regions than along the arms, and star-forming clumps in regions of low velocity dispersion (Sukay et al.,  in prep.).  This unexpected result is entailing new modeling efforts and theoretical interpretations, requiring a re-evaluation of established paradigms of star formation. \\
The kinematic analysis will be used to look for inflowing or outflowing gas and to investigate gas accretion, and ejection processes, and to assess the system gas budget and star formation duration. Finally, we will look for signs of merging activity and interactions. In summary, the kinematic analysis will reveal the origin of the high star formation activity and SMBH growth in the selected targets and the cause of their decline. \\
\textbf{(5) Jet impact:}
To map the effects of the jet on the ISM from the nuclear regions to the galaxy outskirts, we will place multiple IFS to fully cover the nucleus, the galaxy, and the radio structures. The most extended radio structures in the selected targets have sizes $\lesssim$2\,arcsec (see Fig.~\ref{fig:lrad_size}), and four sources contain multiple radio structures (see Radio morphology in Table~\ref{tab:targets}). Thanks to the possibility of placing up to 12 IFS across 24\arcsec\ and next to each other (with only 0.33\arcsec\ gaps in between), VESPER will provide the unique possibility of mapping and obtaining resolved spectra along the full extent of the radio jet.\\ 
\textbf{(6) Environment characterization:}
Since the type of targets we have selected is often associated with overdense environments \citep{ramos_almeida13}, the remaining IFSs will serve to observe nearby galaxies to assess whether the target belongs to an overdensity. They will be also placed on bridges and tidal tails, often present in this type of systems (see e.g., the case illustrated in Fig.~\ref{fig:red_QSO}), to determine the level and type of interaction.  \\
The unique set of VESPER data and the measurements listed above will finally reveal the origin and role of the radio emission, and provide constraints on different feedback modes.

\subsection{Multi-dimensional analysis}

A key goal of the proposed observations is to identify and physically disentangle the multiple kinematic components that coexist in obscured AGN hosts, including rotating disks, AGN- or star formation-driven outflows, inflows, and jet-induced shocks. The SHARP/VESPER data provide the combination of spatial and spectral resolution required to separate these components both spectroscopically and morphologically.
We will perform a resolved analysis of the emission-line profiles (e.g., \Ha, \oiiib, \sii) adopting multi-component Gaussian fitting. The spectral resolution (R $\sim$ 3000) allows us to resolve line profiles into narrow (FWHM $\leq$ 1000\,\kms), intermediate, and broad (FWHM $>$ 2000\,\kms) components, as well as to identify velocity offsets of a few tens of \kms. Each component will be characterized by its velocity centroid, FWHM, and flux, enabling the construction of separate kinematic maps for each component.
The decomposition will be guided by both statistical criteria (e.g., Bayesian Information Criterion to determine the number of required components) and physical consistency across adjacent regions. To ensure robustness, we will consider imposing continuity constraints to have coherent spatial structures.
The identification of the physical origin of each kinematic component will rely on a combination of diagnostics:\\
(i) \textbf{Kinematic signatures:}
Rotating disks will be identified through smooth velocity gradients and low velocity dispersion, while outflows/inflows will appear as red- or blue-shifted components with elevated dispersion. Residual maps (after subtracting rotating disk models) will be used to isolate non-circular motions \citep[see e.g.,][]{tadaki20b}.\\
(ii) \textbf{Ionization diagnostics:}
Emission-line ratios (e.g., BPT diagrams, \oiiib/\hbeta, \nii/\Ha) will be measured separately for each kinematic component to distinguish AGN photoionization, star formation, and shock excitation. \\
(iii) \textbf{Spatial correlation with radio structures:}
By comparing the location and orientation of the kinematic components with high-resolution radio maps, we will directly test whether disturbed kinematics and shocked gas are co-spatial or perpendicular to the radio jet, allowing us to investigate the coupling between the jet and the disk and distinguish jet-driven feedback from radiatively driven winds \citep[see e.g., ][]{mazzalay13}.  \\
(iv) \textbf{Velocity dispersion and energetics:}
The velocity dispersion and the line luminosity for outflowing components will be used to estimate mass outflow rates and kinetic power, enabling a quantitative comparison between different powering sources.\\
This multi-dimensional approach (spectral decomposition, spatial analysis, ionization diagnostics, and radio morphology) will allow us to robustly disentangle overlapping kinematic structures even in heavily obscured systems. Similar analyses applied to JWST and spectral cubes datasets have revealed complex configurations of outflows, shocks, and radio jets (see e.g., Fig. 2, and \citealt{speranza24}), demonstrating the feasibility of this method.
In summary, SHARP will enable a fully resolved separation of the different galactic components, providing a direct view of how jets, winds, and star formation jointly shape the ISM.

\subsection{Overcoming current limitations}

Thanks to the extensive observational effort at radio wavelengths carried out on the selected sample of radio DOGs, the proposed study will not suffer of limitations due to the paucity of available angular scales. These have permitted to establish that the radio emission comes from a relativistic jet, and finally obtain observational constraints on the radio feedback mode at high redshifts.\\
The capabilities of SHARP@ELT will enable to measure the density, and the geometry of outflows instead of making simplified assumptions to compute their energetics \citep{harrison18}. The expected sensitivity will provide enough S/N to measure variations even in faint lines and to examine the spectra of the sub-components in individual systems rather than using stacked spectra \citep{zakamska14}.
The spectral coverage will give access to the rich ISM diagnostic of the rest-frame visible spectrum, thereby enabling the study of the impact and interplay of ouflows and jets in AGN up to $z\sim2.5$.
It is now well established that the environment plays an important role in determining the fate of a galaxy as this regulates the gas accretion flow and the interactions with nearby galaxies. The FOV and multi-IFS capabilities of VESPER will provide the capability of investigating the environment of the selected targets with high efficiency.

\subsection{Depth and exposure time}
\label{sec:etc}
To estimate the depth necessary to carry out the analysis described above we consider both the continuum brightness and the strength of the emission lines.  Since we aim at carrying out resolved SED fitting and detect absorption features in the visible rest-frame, we need a S/N$\sim$10 in regions that match the angular resolution of the resolved radio emission, that is 0.1\arcsec\ (spatial resampling of 3$\times$3 pixels). The K-band magnitudes of the selected targets range from 19.6 to 21.8, with a median value of 20.4. 
To derive the typical exposure time per source, we assume a K-band magnitude of 21 (8 out of the selected 11 are brighter than this limit) and a surface brightness distribution described by a Sersic profile with index $n=1$ and effective radius 350\,mas \citep[$\sim$3\,kpc at $z=2$; ][]{polletta24,vanderwel25}. According to the SHARP Exposure Time Calculator (ETC v0.6\footnote{https:\/\/sharp.lambrate.inaf.it\/}), we will reach a S/N of $\sim$10 in the center and of $\sim$4 at the effective radius  at the reference wavelength (2.2$\mu$m) with an exposure of 5\,hr and with a spatial binning of 3$\times$3 and a spectral binning of two resolution elements (8 pixels). \\
From the estimated SFRs (see Table~\ref{tab:targets}), we derive an \Ha\ luminosity from star formation of $\sim(3-20)\times 10^{43}$\ergs, assuming the SFR--L(\Ha) relation in \citet{kennicutt12} and a Chabrier initial mass function \citep[IMF; ][]{chabrier03}. At the sources redshift these luminosities correspond to total \Ha\ fluxes of $1.7-3.7\times10^{-16}$\ergcm2s, after applying the estimated extinction (\av$\simeq1-3$). With 5\,hr exposure the ETC (v0.6) tells us that we would detect a line with flux of $1.7\times10^{-16}$\ergcm2s, and width of FWHM$\sim$1000\,\kms\ ($\Delta\lambda=55$\,\AA\ at $z=1.53$ and 75\AA\ at $z=2.44$) with S/N$\sim$42 at $\lambda=1.66\mu$m (corresponding to \Ha\ at the sample lowest redshift), and with S/N$\sim$16 at $\lambda=2.26\mu$m (\Ha\ at the highest redshift). These S/N estimates are derived for the central pixel and assuming the same Sersic profile as for the continuum, per resolution element, and with a 3$\times$3 spatial resampling. At the effective radius the predicted S/N are 17, and 6 at the lowest and highest redshift of the sample, respectively. These S/N are sufficient to carry out the spectral and kinematic analysis described in Sect.~\ref{sec:strategy}.\\
In conclusion, an average integration time of 5\,hr per source would produce continuum and line maps with the required S/N to carry out a spatially resolved analysis of the selected targets. Assuming an average number of three VESPER IFS to cover the full extent of a target and its radio jet, we will be able to observe nine additional sources per field, and ten in total, including the selected target. Thus, with a total integration time of 11 radio DOGs $\times$5\,hr\,=\,55\,hr$+$overheads, we will observe, in total, at least 110 sources for which we will be able to carry out the analysis described above.

\section{Synergy with other facilities} 
\label{sec:synergy}
\begin{figure*}%[ht!]
    \centering
    \includegraphics[width=0.95\textwidth]{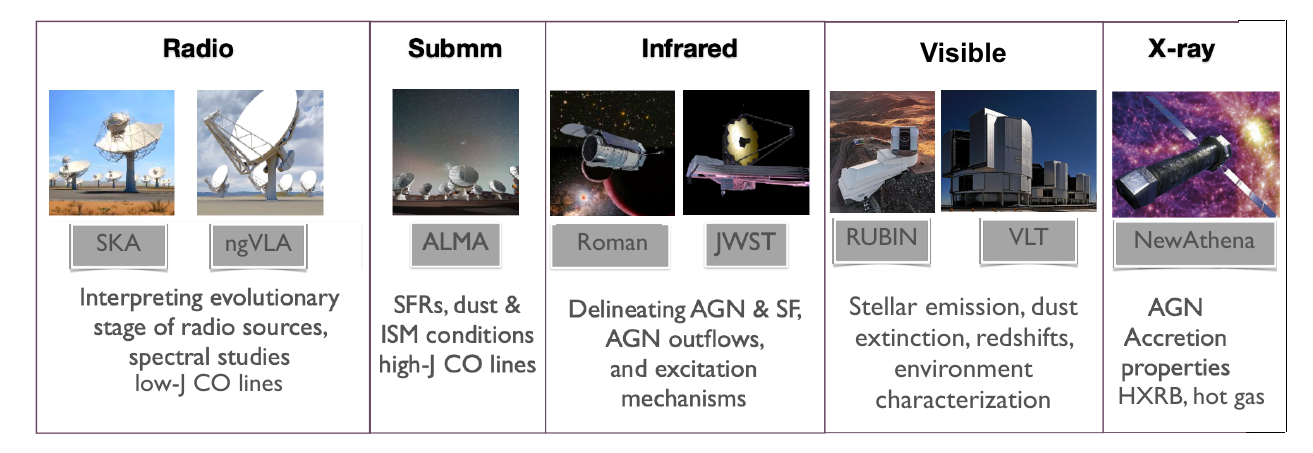}
   \caption{Synergies of the ELT with current and future multi-wavelength facilities.  Some of the scientific questions that will be enabled by these facilities through observations of radio DOGs are reported below the corresponding spectral window.  A complete understanding of young radio AGN in luminous quasars at cosmic noon will greatly benefit from multi-wavelength observations at high spatial and spectral resolution. (Adapted from \citet{patil_phd}).}
    \label{fig:synergies}
\end{figure*}

The proposed study will greatly benefit from other facilities that will be available in the near future. In Fig.~\ref{fig:synergies}, we briefly summarize the main contribution per spectral window, and below we report in more detail some of the key measurements that can be obtained towards a full understanding of AGN feedback at cosmic noon.\\
\textbf{Radio: } The next generation of radio telescopes, such as the SKA and the ngVLA, expected to achieve full science operations by 2028, and 2034, respectively, will
provide additional targets and more details on the radio DOG sample already in hand.  In particular, there will be an improvement in resolution and sensitivity that will permit to search for faint emission that might reveal their past radio activity and provide constraints on the life cycle of the jet. \\
\textbf{(Sub-)Millimeter: } Observations at millimeter wavelengths at high angular resolution, as those carried out by the Atacama Large Millimeter Array (ALMA), can image and resolve the molecular gas distribution, and measure its mass, kinematics, density and temperature. These measurements would constrain the molecular gas fraction, the star formation efficiency and consumption rate.  With these measurements it will be possible to investigate whether the jet or the outflow is destroying the molecular gas or injecting turbulence thus inhibiting star formation. It will be also possible to investigate how the cold gas couple with the ionized outflows and the jet. Finally, they can also identify companion galaxies that are too obscured to measure their redshift from the visible spectrum or to be detected. \\
\textbf{Mid-IR: } The mid-IR window is rich with spectral features that carry information on the warm gas components of the ISM, reveal the most obscured regions, shocked gas and AGN heated dust.  These observables could provide additional constraints on the energetic contribution from star formation, the jet and AGN radiation.  The JWST, and the Nancy Grace Roman Space Telescope (operational in 2027-2032) will be able to provide resolved maps and spectra necessary to study the warm dust in the selected targets.\\
\textbf{Visible: } The SED fitting of the selected targets would benefit from multi-band data at high spatial resolution and at shorter wavelengths than the SHARP window, such as those that will be delivered by the Rubin Vera Observatory. To investigate their global environment, the Multi-object Optical and Near-IR spectrograph (MOONS) at the VLT (starting operations in 2026) would be an efficient machine as it could deliver spectra for $\sim$1000 galaxies within 12\arcmin\ enabling the identification of galaxies at the same redshift.\\
\textbf{X-Ray: } Radio DOGs are also expected to be X-ray luminous. X-ray observations, such as those that will be provided by NewAthena (launch expected in 2037) would be crucial to reveal the most energetic events, determine the AGN accretion power, and search for hot gas.\\
In summary, each spectral window offers a distinct diagnostic power and discovery potential; taken together, they provide a comprehensive view of the past and current activity of these AGN, trace their evolutionary pathways, and uncover the mechanisms driving the Universe’s most active phase and its subsequent decline.
\section*{Acknowledgments}
The SHARP team acknowledges support by Bando Ricerca Fondamentale INAF 2022, Techno-Grant "SHARP" - 1.05.12.02.01, Bando Ricerca Fondamentale INAF 2023 - 1.05.23.04.01, and Bando Ricerca Fondamentale INAF 2024, Large-Grant "SHARP" - 1.05.24.01.01.\\
The National Radio Astronomy Observatory and Green Bank Observatory are facilities of the U.S. National Science Foundation operated under cooperative agreement by Associated Universities, Inc.
Some of the data products presented herein were retrieved from the Dawn JWST Archive (DJA).  DJA is an initiative of the Cosmic Dawn Center (DAWN), which is funded by the Danish National Research Foundation under grant DNRF140.

\printcredits

\bibliographystyle{cas-model2-names}

\end{document}